# Teaching at the Intersection of Social Justice, Ethics, and the ASA Ethical Guidelines for Statistical Practice

Rochelle E. Tractenberg

Collaborative for Research on Outcomes and -Metrics, and
Georgetown University, Building D, Suite 207, 4000 Reservoir Road NW, Washington, DC 20057

Acknowledgement: There are no actual or potential conflicts of interest. Opinions expressed in this article are the authors' own.

Running Head: Teaching social justice with the Ethical Guidelines for Statistical Practice

To appear (verbatim) in: *Proceedings of the 2023 Joint Statistics Meetings, Toronto, Ontario, Canada*. Alexandria, VA: American Statistical Association
Preprint Available on StatArXiv: TBN

Abstract

Case studies are typically used to teach "ethics", but emphasize narrative. When the content of a course is focused on formulae and proofs, it can seem divergent or distracting – for both instructor and learner - to introduce a case analysis and the knowledge, skills, and abilities required for effective teaching and learning. Moreover, case analyses are typically focused on issues relating to people as data donors ("human subjects"): obtaining consent, dealing with research team members, and/or potential institutional policy violations.While relevant to some research, not all students in quantitative courses plan to become researchers, and ethical practice – of mathematics, statistics, data science, and computing – is an essential topic regardless of whether or not the learner intends to do research, or use human subjects data. It is a mistake to treat "training in ethical practice" and "training in responsible conduct of research" as the same thing, just as incorrect as it is to assume that "training in ethical practice" is the same irrespective of what the learner will actually be practicing. By the same token, it is also a mistake to treat "social justice" as a proxy for "ethical practice"; however, the topic of social justice may seem much more accessible to students, and more interesting to instructors. This paper offers concrete recommendations for integrating social justice content into quantitative courses in ways that limit the burden of new knowledge, skills, and abilities for instructors but also create reproducible and actionable assessments. Five tools can be utilized repeatedly to integrate social justice into a course in a way that (also) meets national calls to integrate "ethics"; minimizes the burden on instructors to create and grade new materials and assignments; minimizes the burden on learners to develop a sometimes unexpected skill set to complete a case analysis; and maximizes the likelihood that the ethics content will be embedded in the learners' cognitive representation of the statistical and data science knowledge being taught in the same course. These tools are:

    a. Curriculum Development Guidelines
    b. 7-task Statistics and Data Science Pipeline
    c. ASA Ethical Guidelines for Statistical Practice
    d. Stakeholder Analysis
    e. 6-step Ethical Reasoning paradigm

This paper discusses how to introduce ethical reasoning (e), stakeholder analysis (d), and the ASA ethical practice standards (c) authentically in quantitative courses, using the Statistics and Data Science Pipeline (b) and internationally recognized guidelines for effective higher education curriculum development (a). The tools and frameworks also offer structure for teaching with or about "social justice", and facilitate ensuring that changes made to any course are evaluable and generate actionable assessments for learners.

## 1. Introduction

Ethical practice of mathematics, statistics, data science, and computing is an essential topic regardless of whether or not the learner intends to do research, or use human subjects data. However, the dominant paradigm for preparing ethical practitioners in these quantitative areas, or even those who would utilize mathematics, statistics, data science, and computing in their practice of other jobs, professions, and roles, is highly limited. Firstly, "training" is primarily focused on ensuring respectful treatment of human "research" subjects. Secondly, training tends to promote the idea that every discipline has its own specific ethical norms - so the training is intended to not be very specific. Thirdly, training is often limited to only those who engage in research or who are on a path committed to research in their future job. These three factors are severely impacting perceptions of utility of "ethics training" by instructors and also by learners in higher education. Meanwhile, the importance of "ethics" in the training of future practitioners in -or with - mathematics, statistics, data science, and computing has been explicitly featured by the American Statistical Association (ASA 2014) and the National Academies (NASEM 2018) for statistics and data science, respectively.

Clearly, for mathematics, statistics, data science, and computing, the ethical treatment of human subjects or data donors is an unnecessarily narrow scope of ethical practice. Professional practice standards for computing (ACM 2018) and statistics and data science (ASA 2022) highlight the range of behaviors where ethical considerations are relevant. Emerging ethical practice standards for mathematics do, as well (Tractenberg et al. 2023). While ACM and ASA have their own ethical practice standards, these overlap to a great degree (Tractenberg, 2020; 2022-A; 2022-B) -particularly when considering computationally intensive or dependent statistics, and data science. The emerging ethical practice guidelines for Mathematics are strongly aligned with those of ACM and ASA (Tractenber, Piercey & Buell 2023). Park & Tractenberg (2023) and Tractenberg & Park (2023) show strong alignment between the ASA Ethical Guidelines for Statistical Practice and guidance for both US and international official statistics. Rather than support the notion that each discipline has its own and varying ethical norms, these recent findings suggest that these disciplins share a great deal in terms of what constitutes "ethical practice" or use of the discipline or its tools in the workforce (see also Tractenberg, Lalonde & Thornton, 2023).

Finally, the idea that only researchers need training in ethical practice is untenable and only serves to undermine the legitimate interests in promoting ethical use of, contributions to, and practice in, each of these disciplines (see Tractenberg, Piercey, & Buell, 2023). Therefore, this paper is intended to engage instructors who are interested in integrating ethics content and/or the construct of "social justice" into their teaching of mathematics, statistics, data science, and computing. For those specifically interested in social justice, the intention is to promote the creation of learning experiences that will result in the same level of learning and accomplishment of learning outcomes irrespective of any time-delimited or specific social justice topic. These tools and techniques to be discussed are conceptualized (by the author, an instructor in higher education & cognitive scientist) for use in higher, graduate, and post-graduate education; however, they could easily be adapted for use in high school settings.

There are tools that can be utilized repeatedly to integrate social justice into a course in a way that (also) meets national calls to integrate "ethics" across the statistics (ASA 2014) and data science (National Academies, 2018) curriculum; minimizes the burden on instructors to create and grade new materials and assignments; minimizes the burden on learners to develop an unexpected skill set to complete a case analysis; and maximizes the likelihood that the ethics content will be embedded in the learners' cognitive representation of the statistical and data science knowledge being taught in the same course. These tools are:

a. Curriculum Development Guidelines (Tractenberg et al. 2020);
b. the seven -task statistics and data science pipeline (Tractenberg 2022-A; 2022-B);
c. the ASA Ethical Guidelines for Statistical Practice (ASA, 2022);
d. the Stakeholder Analysis (Tractenberg, 2019);
e. the 6-step Ethical Reasoning paradigm (Tractenberg & FitzGerald, 2012; Tractenberg 2022-A; 2022-B)

In the following sections, each of these tools is briefly introduced and discussed.

## 2. Curriculum Development Guidelines

Whenever contemplating or executing curriculum development, evaluation, or revision, instructors should follow accepted Guidelines for Instruction and Curriculum Development (e.g., Tractenberg et al. 2020). This is true whether the innovation is an entire program, a single course, or materials for an assignment. Specifically instructors should,

A. Identify and follow a formal paradigm for curriculum or instructional design.
B. Focus on Learning Outcomes (**LOs**) first, to inform all other decisions about the curriculum and instruction.
  B1: Leverage LOs to explore and identify appropriate **learning experiences (teaching/class activities)**.
  B2: Leverage LOs to select **content** that is appropriate for the learning experiences and promotes the LOs.
  B3: **Assess** learning based on achievement of LOs using formative and summative assessment, as appropriate.
C: Plan and execute an actionable evaluation of the curriculum and instruction.
D: Document and share the features of the curriculum or instruction – including criteria for their success – with learners.

The rationale and background, as well as further reading, for these Guidelines are discussed in Tractenberg et al. (2020) and Tractenberg (2022-C).

## 3. The Statistics and Data Science Pipeline

When contemplating new course material or new knowledge, skills, or abilities to integrate into courses on statistics and data science, rather than follow the topical organization of the typical textbook, it can increase retention to incorporate authentic organization (see Tractenberg, 2023-A). Seven tasks fall along a general pipeline that matches the scientific method (Tractenberg 2022-A; 2022-B), and so can easily be mapped to any syllabus or course outline for statistics and data science. These tasks can serve to highlight the relevance of ethical practice standards throughout a project **and a career.**

The seven-task structure also provides some "lexical grounding" for learners; that is, embedding ethical content within the instruction or lessons on each task encourages learners to map the new information (practical task plus the ethical aspects of executing that task) to developing knowledge (early in their training) or to existing knowledge later in training or when they enter practice. The seven tasks are:

1. plan/design
2. collect/munge/wrangle data
3. analysis – run or program to run
4. interpret
5. report & communicate
6. document your work

7. work on a team

Considering ethical practice at each of these steps along the SDS pipeline can bring structure to both instruction and also to how students organize the information about ethical practice, at each step. No matter where they enter -or exit- this pipeline, there are responsibilities, and opportunities, to accomplish each step ethically. Moreover, it is essential for learners as well as instructors and practitioners to recognize that there are **two** distinct dimensions to ethical practice of statistics and data science (Tractenberg 2022-A; 2022-B). The two dimensions are:

**Dimension 1**: To practice ethically, i.e., to execute each task in accordance with the Ethical Guidelines for Statistical Practice (ASA, 2022).
**Dimension 2**: To identify, and respond to, unethical actions/requests, i.e., recognize unethical behavior and act to resolve or correct it. This can be done with the Ethical Guidelines for Statistical Practice, but might also require consultation with local rules, regulations, policies, and laws.

Instruction should focus first on the ethical *practice* dimension (Dimension 1): how to utilize the ASA Ethical Guidelines for Statistical Practice (see next Section) to execute all statistical work - i.e., in each of the tasks on the SDS pipeline (see e.g., Tractenberg, Lalonde, & Thornton, 2023). Statistical practice, as defined in the Guidelines, "includes activities such as: designing the collection of, summarizing, processing, analyzing, interpreting, or presenting, data; as well as model or algorithm development and deployment". (ASA 2022) This definition touches on each of the pipeline tasks, implicitly including working on a team through presentation and communication. This dimension, simply practicing ethically, will hopefully comprise the vast majority of learners' eventual statistical and data science practice, and this is also a simpler dimension to teach how to do ethically. This is, unfortunately, almost never considered in "ethics training", and is definitely not how case analyses are used to teach "ethics". When considering social justice, it is this dimension that can facilitate learners' engagement with the ideas that the way they execute their typical tasks and work (ethically!) will be an important tool in their activities to promote social justice.

It is important for learners to recognize that they can serve social justice by practicing ethically - and specifically following the ASA Ethical Guidelines for Statistical Practice (ASA, 2022). But if they ever encounter a situation where statistical practices are, or are under pressure to be, misused or misrepresented, there are specific obligations in the Ethical Guidelines for those situations, and later in this paper, a concrete method for identifying and responding to such situations is discussed. The method can be used in any context involving ethical practice, including situations involving social justice.

The SDS Pipeline can be effectively utilized to teach ethical practice along each of these dimensions (Tractenberg, 2022-A). It is essential to recognize that simple awareness, or memorization, of the ASA Ethical Guidelines for Statistical Practice is not sufficient to support either of these dimensions.

## 4. The ASA Ethical Guidelines for Statistical Practice

The ASA Ethical Guidelines for Statistical Practice (ASA 2022) are an ethical practice standard that are intended specifically to promote ethical decision making by statistical practitioners irrespective of job title, level, or field of degree. While the SDS Pipeline can reinforce the relevance and importance of both ethical statistical practice and how thoroughly the responsibilities to execute each task ethically permeate through work and workflows, the ASA Ethical Guidelines reinforce the idea that to be considered an ethical statistics practitioner - no matter what your actual job title, or how frequently or infrequently you engage in actual statistical practice - all of the relevant Guideline elements and Principles should be applied and/or followed.

**Table 1:** ASA Ethical Guidelines for Statistical Practice (ASA 2022): Principles (number of elements)

**A. Professional Integrity & Accountability**: Professional integrity and accountability require taking responsibility for one's work. Ethical statistical practice supports valid and prudent decision making with appropriate methodology. The ethical statistical practitioner represents their capabilities and activities honestly, and treats others with respect. (12 elements)

**B. Integrity of data and methods:** The ethical statistical practitioner seeks to understand and mitigate known or suspected limitations, defects, or biases in the data or methods and communicates potential impacts on the interpretation, conclusions, recommendations, decisions, or other results of statistical practices. (7 elements)

**C. Responsibilities to Stakeholders**: Those who fund, contribute to, use, or are affected by statistical practices are considered stakeholders. The ethical statistical practitioner respects the interests of stakeholders while practicing in compliance with these Guidelines. (8 elements)

**D. Responsibilities to research subjects, data subjects, or those directly affected by statistical practices**: The ethical statistical practitioner does not misuse or condone the misuse of data. They protect and respect the rights and interests of human and animal subjects. These responsibilities extend to those who will be directly affected by statistical practices. (11 elements)

**E. Responsibilities to members of multidisciplinary teams**: Statistical practice is often conducted in teams made up of professionals with different professional standards. The statistical practitioner must know how to work ethically in this environment. (4 elements)

**F. Responsibilities to Fellow Statistical Practitioners and the Profession**: Statistical practices occur in a wide range of contexts. Irrespective of job title and training, those who practice statistics have a responsibility to treat statistical practitioners, and the profession, with respect. Responsibilities to other practitioners and the profession include honest communication and engagement that can strengthen the work of others and the profession. (5 elements)

**G. Responsibilities of Leaders, Supervisors, and Mentors in Statistical Practice**: Statistical practitioners leading, supervising, and/or mentoring people in statistical practice have specific obligations to follow and promote these Ethical Guidelines. Their support for – and insistence on – ethical statistical practice are essential for the integrity of the practice and profession of statistics as well as the practitioners themselves. (5 elements)

**H. Responsibilities regarding potential misconduct**: The ethical statistical practitioner understands that questions may arise concerning potential misconduct related to statistical, scientific, or professional practice. At times, a practitioner may accuse someone of misconduct, or be accused by others. At other times, a practitioner may be involved in the investigation of others' behavior. Allegations of misconduct may arise within different institutions with different standards and potentially different outcomes. The elements that follow relate specifically to allegations of statistical, scientific, and professional misconduct. (8 elements)

**APPENDIX: Responsibilities of organizations/institutions:** Whenever organizations and institutions design the collection of, summarize, process, analyze, interpret, or present, data; or develop and/or deploy models or algorithms, they have responsibilities to use statistical practice in ways that are consistent with these Guidelines, as well as promote ethical statistical practice. (Organizations 7 elements; Leaders 5 elements; 12 elements total)

In addition to their applicability to anyone, at any career stage, and with any job title or training who utilizes statistical practices, the ASA Ethical Guidelines also pertain *whenever individuals engage in statistical practice*, which includes designing the collection of, summarizing, processing, analyzing, interpreting, or presenting, data; as well as model or algorithm development and deployment. Students may be surprised to learn that organizations and institutions that hire or employ statistical practitioners also have responsibilities to use the outputs of statistical practices ethically, and to ensure that statistics practitioners are allowed to, and supported when they, strive to follow the ASA Ethical Guidelines.

The ASA Ethical Guidelines are a core tool for teaching ethical statistical practice in each dimension because they represent a concrete ethical practice standard (see Gillikin, 2017; Tractenberg, 2022-A; 2022-B; 2023-A; Tractenberg & Park, 2023) for statistics and data science - that is, these are the description of "ethical practice" - by anyone who uses- or employs people who use- statistical practice as defined above.

Ethical statistical and data science practice is not abstract. It is observable: specifically, the ASA Ethical Guidelines describe what the ethical statistical practitioner does, and looks like while doing it. Students can utilize this knowledge to identify workplaces that share a commitment to ethical practices as well as to social justice or other causes they support.

Federal (US) funders created a "required topics" list, and many institutions that receive federal funding created "training" that meets this requirement. However, "this requirement" is simply - and is implemented simply as if - students hear about these topics. Unfortunately, **there is no requirement from any US funder that the training to become a researcher also trains them to become ethical practitioners of statistics and data science or researchers who use statistics and data science.** Not surprisingly, this topics list-based training requirement has had no apparent effect on ethical research practices – they continue to increase in prevalence (National Academies, 2017). The ASA Ethical Guidelines for Statistical Practice meet and can exceed this requirement - while also introducing statistics and data science practitioners, irrespective of their degree or job title, to what "ethical practice" is and looks like (Tractenberg 2022-A; 2022-B).

### 5. The Stakeholder Analysis

A Stakeholder Analysis (Tractenberg 2019; Tractenberg 2023-B) is a concrete way for learners to begin to perceive the effects of their decisions about statistics and data science (e.g., which method to use, which data to collect or obtain, how to communicate results, etc.) on others, i.e., on stakeholders. In class discussions and written work can ask students to consider *or list* harms and benefits that will happen given an action or decision. An empty stakeholder analysis (SHA) table appears below.

**Table 2:** Stakeholder analysis (Reprinted from Tractenberg, 2019)

| Potential result: | HARM[4,5] | BENEFIT[4,5] |
|---|---|---|
| Stakeholder[1]: | | |
| YOU[2,3] | | |
| Your boss/client | | |
| Unknown individuals[2] | | |
| Employer | | |
| Colleagues | | |
| Profession | | |
| Public/public trust | | |

Notes to Table:
1 Learning how to use this table and complete a case analysis is essential for enabling informed decisions about ethical challenges.
2 SHA permits definition of "social justice" in terms of (im)balance of harms/benefits, or any other ratio.

3 Recognize whether or not benefits or harms accrue to any stakeholders. Are some harms "worse" or some benefits  "better"? Does this add/decrement ethical aspect of act/decision?
4 All harms are not the same; all the benefits are not the same; and harms and benefits are not exchangeable.
5 Knowing to whom harms may accrue can point to where the ASA Guidelines can assist in decision making

This table is taken from Tractenberg, 2019. An in-class activity is, for any decision or circumstance (e.g., choice of method or assumptions), to fill in the table to identify & quantify what is socially just/unjust. For any situation or circumstance, "injustice" can be defined based on the SHA- e.g., whenever the public trust has harms or potential harms identified, and no benefits, or, when an employer would receive a benefit but the public, or unknown individuals, would accrue harms. The notes in the table above can be used to structure the learning outcomes for integrating "social justice" or ethics content into a course. The table below shows how Messick's three questions for valid assessment (Messick 1994; adapted in Tractenberg, 2023-B) can be answered with respect to the SHA and/or the ASA Ethical Guidelines.

**Table 3:** Stakeholder analysis using the ASA Ethical Guidelines can generate repeatable, gradable learning outcomes -irrespective of definition of "social justice".

| Three critical questions (Tractenberg, 2023-B) | Learning outcomes from Stakeholder Analysis |
|---|---|
| **What are the knowledge, skills, and abilities (KSAs) the course (or curriculum) should lead to?** | 1. Describe how different individuals ("stakeholders") may be affected by decisions and actions;<br>2. Enumerate harms and benefits that are most clearly relevant for each stakeholder with respect to the activity; and<br>3. Identify which **ASA Ethical Guideline** or policy principles, practices, and/or specific elements seem most relevant to the specific decision/action. |
| **What actions/ behaviours by the students will reveal these KSAs?** | Depends on Bloom's level target: multiple-choice; matching; fill in the blank; list; written answer/essay |
| **What tasks will elicit these specific actions or behaviors (reveal whether students did learn the target KSAs)?** | General/re-usable: Stakeholder analysis of harms/benefits that will occur if: **assumptions** don't hold; **approximations** are incorrect; or **applications** are inappropriate. |

Stakeholder analysis allows the instructor to reproducibly, and evaluably, map social justice onto ethical practice of statistics and data science. Instructors can even ask students to define "social injustice" for themselves, and then apply that to the tasks associated with the Stakeholder Analysis. Moreover, considerations of stakeholders can be integrated in more quantitative courses whenever assumptions, approximations, or applications are mentioned (Tractenberg, 2023-B).

## 6. Ethical Reasoning Paradigm

"**Ethics is the effort to guide one's conduct with careful reasoning**. One cannot simply claim "X is wrong."; Rather, one needs to claim "X is wrong because (fill in the blank)"." (Briggle & Mitcham, 2012, p. 38). (emphasis added).

Ethical Reasoning is a learnable, improvable, set of knowledge, skills, and abilities (KSAs; Tractenberg & FitzGerald 2012; Tractenberg 2022-A; 2022-B), and these are:
1. Determining your prerequisite knowledge;
2. Identification of decision-making frameworks;
3. Recognizing an ethical issue;
4. Identification and evaluation of alternative actions;
5. Making and justifying decisions;

6. Reflecting on the decision.

Typical classes on "ethics" or "responsible conduct of research" feature case analysis - which requires all six of these steps to be taught, practiced with feedback, and demonstrated. However, the KSAs of ethical reasoning -which are learnable and improvable (Tractenberg & FitzGerald 2012) are rarely taught, but are most often assumed to be available to students. Note that Briggle & Mitcham (2012) articulate that, "careful reasoning" about conduct is required for ethical behavior. However, this is not a complete or accurate perception of ethical statistical practice. In fact, statistical practitioners (irrespective of level of training or job title) spend the majority of their time on KSA #1 - using the ASA Ethical Guidelines for Statistical Practice (and/or the ACM Code of Ethics) to simply practice ethically. There is a lot to "know", and that prerequisite knowledge is important for ~90% of your work. Thus, Dimension 1 is essential for general, ethical practice in statistics and data science in every task along the SDS pipeline. Instructors can utilize the Stakeholder Analysis table and the ASA Ethical Guidelines along the SDS pipeline to teach, give practice with, and encourage students to demonstrate proficiency with KSA #1. This would go far in encouraging ethical practice of statistics and data science - much further than a focus on the topics of "responsible conduct of research" ever could.

The second dimension of "ethical statistical practice" is considerably more complex to teach and assess effectively, and to demonstrate - both by instructors and by students. However, sometimes *others* make decisions that require us to respond and in those cases, **our** ethical practice may not be sufficient. **We need a method to <u>respond</u>: making and justifying *ethical* decisions**. This decision making requires all 6 of the KSAs of ethical reasoning (Tractenberg 2022-A; 2022-B). As can be seen from the list of KSAs, knowing and following the Ethical Guidelines, with or without the SHA, is not enough to identify and respond to unethical behavior (Dimension 2). The figure below shows how the engagement with higher Bloom's levels of complexity are required for tasks like case analysis with justification of choices (Bloom's level 4), synthesis of disparate sources of information for an argument or rationale (Bloom's level 5), or the execution of judgment or evaluation of justification (Bloom's level 6). Only when a practitioner can reliably engage these higher levels of cognitive complexity in a case analysis can they be fully competent at ethical reasoning through a case analysis (Dimension 2).

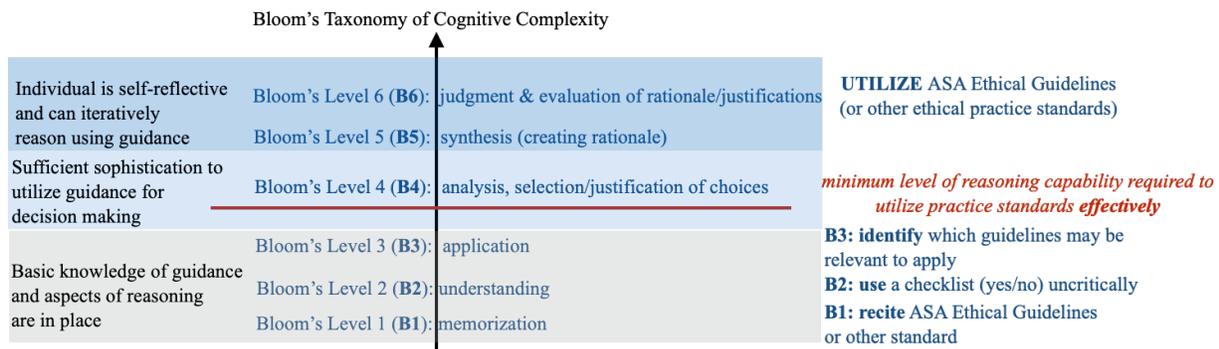

**Figure 1:** Bloom's Taxonomy of Cognitive Behaviors (complexity) required to utilize the ASA Ethical Guidelines, ACM Code of Ethics, the Data Science Ethics Framework, or other standard.

As can be seen in Figure 1, Bloom's level 4 is the minimum level of reasoning that is required to utilize the ASA Ethical Guidelines for Statistical Practice as the ethical practice standard that it is - to guide practice (Dimension 1) and also identify and make a defensible decision about unethical behavior (Dimension 2). Note also that, in the real world (i.e., after the course or program where ethical reasoning was learned and practiced), the individual will be fulfilling both Dimensions 1 and 2 on their own, without the oversight of the instructor. That independent capability is critically dependent on Bloom's level 6 - the ability to judge and evaluate the arguments and behaviors of others. Again, an instructor who focuses on getting students to Bloom's level 4 can do so with a Stakeholder Analysis, where students demonstrate their functional level on Bloom's Taxonomy of Cognitive Complexity using short answer, matching, and other reproducible tasks – even if the content and/or definition of "social justice" varies over time or across students.

## 7. Synthesizing these tools to integrate ethics and social justice

As is discussed in Tractenberg (2023-B), the SHA and case analysis are two tasks that can be used/re-used throughout a course or program to integrate ASA Ethical Guidelines, Stakeholder Analysis, and social justice with quantitative material. Note that tasks and assignments should be designed specifically to provide instruction, practice, and opportunities to demonstrate knowledge, skills, and abilities that are taught and intentionally introduced into a course. Students will rightfully become frustrated if instructors assign a case analysis, but do not provide instruction or practice with doing this complicated type of analysis. The SDS Pipeline can be leveraged to give multiple exposures to either SHA or case analysis throughout a term or program; student responses can be expected to increase in sophistication - using Bloom's Taxonomy of Cognitive Behaviors (Bloom 1956) over time by structuring low, medium, and high cognitive complexity versions of the same task (see Tractenberg, 2023-B).

### 7.1 Dimension 1: practice ethically

Instructors can add tasks involving a Stakeholder Analysis of harms/benefits that will occur: if **assumptions** don't hold; **approximations** are incorrect; or **applications** are inappropriate. Throughout a course with quantitative content, once the Stakeholder Analysis construct is introduced, this can be interjected throughout the course or program to encourage both a deeper and more authentic engagement with the content itself, and also with what their decisions can mean to others beyond them or their work. Deepening student awareness of the practical applicability of the content will increase the strength of its representation in their mental schema for the topic and potentiate its transfer to any other application of the content (Tractenberg, 2022-C). The explicit linkage of the quantitative content to "others" through the Stakeholder Analysis also offers a straightforward way to introduce ethics and social justice into a course without depending on individualized and/or time dependent definitions of these constructs.

Once students are familiar with the Stakeholder Analysis, instructors can then explicitly add the 'ethical practice' dimension (Dimension 1) by asking which ASA Ethical Guideline elements offer guidance if **assumptions** don't hold; **approximations** are incorrect; or **applications** are inappropriate (Tractenberg 2023-B). Recognizing that making assumptions, using approximations, and selecting methods to apply are all *decisions* that practitioners must make can be easily integrated into any quantitative course. This enables instructors to move students towards Bloom's level 4 - the minimum required for effective utilization of an ethical practice standard like the ASA Ethical Guidelines for Statistical Practice (Figure 1). Note that the focus is on ethical practice; instructors who want to emphasize social justice can do so from within the ASA Ethical Guidelines while ensuring that -even if there is no chance of social *injustice* - students will be learning how to be ethical practitioners of statistics and data science. Importantly, a focus solely on "social justice" does not offer the same benefit because ethical statistical and data science practice is not necessarily a component of social justice, which may depend on time, culture, and other situational factors.

### 7.2 Dimension 2: Identify & respond to unethical act/request

As noted earlier, the second dimension of "ethical statistical and data science practice" is considerably more complicated to teach and assess than the first one is. The Figure below shows how the Ethical Reasoning KSAs and case analysis depend on Bloom's cognitive complexity.

Case analysis

Stakeholder analysis (and other tasks like matching, short answer, fill in the blank)

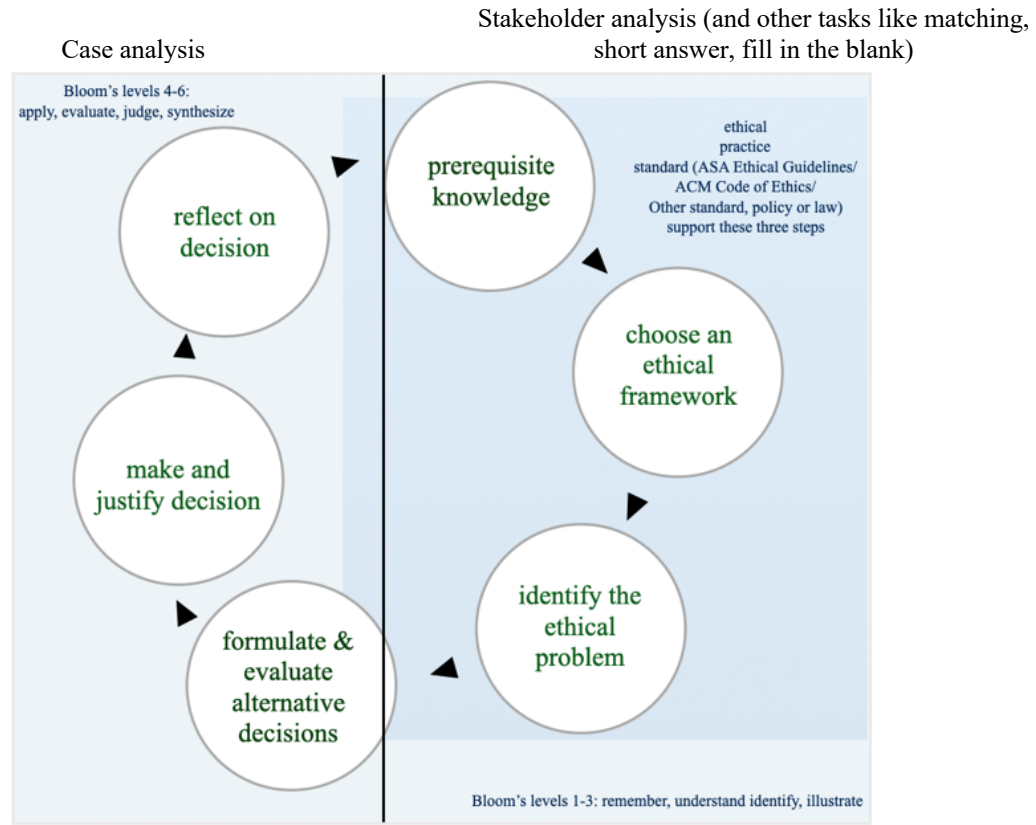

**Figure 2:** The six elements (knowledge, skills, abilities or KSAs) of Ethical Reasoning mapped with Bloom's Taxonomy cognitive complexity requirements to utilize the ASA Ethical Guidelines for Statistical Practice, ACM Code of Ethics, or other reference. Case analysis requires advanced Bloom's cognitive capabilities (levels 4-6) (Dimension 2), whereas lower Bloom's complexity activities are needed for stakeholder analysis and other tasks that support ethical practice (Dimension 1).

Once students buy into the idea of the Stakeholder Analysis, and the facts that they make decisions throughout the SDS Pipeline and that those decisions can affect stakeholders differentially and not-exchangeably, the extension of instruction beyond basic Bloom's levels of cognitive complexity may be more interesting to students. Encouraging students to recognize how their decisions affect others (Dimension 1 and Stakeholder Analysis) can help them buy into the challenges of thinking harder, and with greater sophistication, about topics like 'social justice'. Buy-in to aspects of ethical practice of statistics and data science by students is important because of well-documented resistance by students in higher education to new or different ways of thinking (Cavanagh et al. 2016). Thus, the elements outlined in this paper can all be utilized to build up students' schemas for ethical practice of statistics and data science while enabling instructors to engage with context- and time-dependent aspects of social justice. Dimension 1 will be more straightforward to teach and assess student learning, and so is an ideal starting point for integrating ethical content and aspects of social justice into any quantitative course or program. It is essential to recognize that Dimension 1 will be critical foundation for ethical practice of statistics and data science; and that teaching and learning along Dimension 2 will be facilitated by a prior focus on Dimension 1. Sharing instructor logic in terms of structuring content, practice, and assessment with the two distinct dimensions, and the ways in which learner sophistication is expected to develop, can also increase student buy-in and active engagement in their own growth and development.

## 8. Discussion and Conclusions

Ethical reasoning is a way of thinking that requires the individual to assess what they know about a potential ethical problem – their prerequisite knowledge, and in some cases, how behaviors they observe, are directed to perform, or have performed, diverge from what they know to be ethical behavior. Ethical reasoning is a learnable, improvable set of knowledge, skills, and abilities that enable learners to recognize what they do and do not know about what constitutes "ethical practice" of a discipline, and in some cases, to contemplate alternative decisions about how to first recognize, and then proceed past, or respond to, such divergences. A stakeholder analysis is part of prerequisite knowledge, and can be used whether there is or is not an actual case or behavior/situation to react to -i.e., for both Dimensions 1 and 2 of ethical practice of statistics and data science. When teaching courses with primarily quantitative content, a stakeholder analysis is a useful tool for instruction and assessment of learning: it can be used to both integrate authentic ethical content and encourage careful quantitative thought. This paper discusses how to introduce ethical reasoning, stakeholder analysis, and ethical practice standards authentically in quantitative courses. Specifically, using published frameworks and tools to get social justice themes into a quantitative course without overwhelming the instructor or the learner, nor creating novelty for students that make the workload seem daunting. The tools and frameworks also offer structure, and facilitate ensuring that changes made to any course are evaluable (did this work?) and generate actionable assessments for learners (did they learn what was intended?). Ethical reasoning leads to evidence-based decision making in the workplace - and part of the evidence is the ASA Ethical Guidelines for Statistical Practice (ASA 2022). When instructors focus instead on "social justice", the evidence in their decision making might be more abstract or otherwise difficult to identify, explicate, and replicate in consistent teaching and assessing. In a formal case analysis that features ethical reasoning, a defensible decision – that can be revisited thoughtfully – is a meaningful dimension of professional identity/professional development. This can contribute an important aspect of buy-in from students and possibly also departments/peer instructors, particularly for those teaching within statistics and data science programs. Ethical reasoning with the ASA Ethical Guidelines facilitates conversations, and promotes justifiable and transparent decisions. This paradigm also easily accommodates time- and context-dependent definitions of social justice (and injustice), while encouraging transparent and actionable (assessable) decisions by students.